\documentstyle[12pt]{article}
\include{amssymb}
\begin{document}
\newcommand{\be}{\begin{equation}}
\newcommand{\ee}{\end{equation}}
\title{On the extension of the Obukhov theorem in non-Riemannian gravity I}
\author{R. Scipioni}
\maketitle
Department of Physics and Astronomy, The University of British Columbia,\\
6224 Agricultural Road, Vancouver, B.C., Canada V6T 1Z1 \footnote{scipioni@physics.ubc.ca}
\bigskip
\bigskip
\bigskip
\bigskip
\bigskip
\begin{abstract}
This is a first paper of a series in which we give some generalizations of the Obukhov theorem in the Tucker-Wang approach to Metric-Affine gravity in which we consider more general actions containing scalar and in general fields which do not depend on the metric or connection.\\
\\
04.20.-q, 04.40.-b, 04.50.+h, 04.62.+v
\end{abstract}
\newpage
\section{INTRODUCTION}
Recently much effort has been devoted to the study of non-standard gravitational theories that is theories which allow for nonmetricity and torsion of spacetime. The usual approach to this generalization of Einstein theory goes through the gauge field method which permits to obtain a gauge theory of gravity starting from the affine group $A(n,R)$ [1].\\
In this approach the metric $g_{ab}$, the connection $\omega^{a}{}_{b}$ and the coframe $e^{a}$ are considered as three independent gauge potentials whose fields are the non-metricity $Q_{ab}$ the curvature $R^{a}{}_{b}$ and the torsion respectively $T^{a}$.\\
However when a detailed study is performed we note that the different equations we get are not independent, in particular the one for the coframe and the one for the metric are related, meaning that the approach contains a kind of redundancy [1].\\
It has been suggested by Tucker-Wang [2,3] to drop one of the potentials like the metric or the coframe and use only the connection and the metric, or the coframe as independent variables in a pure variational approach. This approach is motivated also by the fact that when we describe the symmetry reduction from the general affine group, to the group describing the low energy limit of gravity we still have the freedom of choosing the coframe. This permits us to choose an orthonormal coframe and by doing so, the degree of freedom of the metric and the coframe becomes equivalent. The metric can be written as: $g = \eta_{ab} e^{a} \otimes e^{b}$ with $\eta_{ab} = diag(-1,1,1,1,...)$. \\
Recently this approach has been used to prove that a remarkable reduction property occurs in the field equations of certain non-Riemannian models of gravity [4]. This property of reduction has been proved in Metric-Affine gravity too [5] and it is now known as the Obukhov theorem [6].\\
In this paper a generalization of this theorem is given obtained by analysing the properties of the Cartan equation obtained from the connection variation. We stress the different role of its trace part compared with the traceless part. While the latter is used to obtain the non-Riemannian part of the connection as a function of the fields appearing in the action, the former gives a certain relation for the Weyl 1-form Q which in the case of the Obukhov theorem is a Proca type equation.\\
We will show that by considering a more general action we can obtain a more general equation for $Q$ but the functional dependence of $\lambda^{a}{}_{b}$ (the non-Riemannian part of connection 1-forms) as a function of the fields $Q, T, \psi,...$ is the same. The result being that the reduction can then be extended to these more general actions like non-Riemannian scalar gravity or theories in which non-metricity or torsions are coupled to metric invariant terms like electromagnetic fields. This property of generalization has been used recently to exhibit a Black-hole Dilaton solution with non-metricity and torsion [7].\\
We consider also a simple generalization of the Obukhov theorem to the case where the Einstein-Hilbert term is coupled to the Dilaton field.\\
To establish the notation, we use a non riemannian geometry which is specified by a metric tensor field $\bf g$ and a linear connection $\bf \nabla$. using a local coframe $e^a$ with its dual frame $X_{b}$ such that $e^a(X_{b}) = \delta^a_b$, the connections 1-forms satisfy $\omega^{c}{}_{b}(X_{a}) \equiv e^c (\nabla_{X_{a}} X_{b})$. The tensor ${\bf S = \nabla g}$ defines the non metricity of the theory; in a local orthonormal frame the metric tensor is ${\bf g} = \eta_{ab} e^a \otimes e^b$, $(\eta_{ab} = diag(-1,1,1,1,...))$ The non metricity 1-forms are defined by $Q_{ab} \equiv {\bf S}(-,X_{a},X_{b})$ and the torsion 2-forms $T^a \equiv de^a + \omega^{a}{}_{b} \wedge e^b$ the curvature two forms are $R^{a}{}_{b} \equiv d\omega^{a}{}_{b} + \omega^{a}{}_{c} \wedge \omega^{c}{}_{b}$, while the general curvature scalar $R$ is given by $ R \star 1 = R^{a}{}_{b} \wedge \star (e_{a} \wedge e^b)$ in terms of the Hodge operator of the metric.\\
Before going to consider the different cases let us briefly remind what is the main conclusion of the Obukhov theorem.\\
\\
{\bf Obukhov's Theorem}[4-6]:\\
\\
\emph{The field Equations of a general non-Riemannian model of gravity can be reduced to the field equations of an effective Proca-Einstein theory in which the Weyl 1-form $Q$ represents the Proca field.}
\\
\section{Scalar field theories}
\bigskip
To begin our analysis let us consider the action:
\begin{eqnarray}
\int \Lambda [e,\omega] = \int [k R \star 1 + \frac{\alpha}{2} f_{1}(\psi)(dQ \wedge \star dQ) + \frac{\beta_{0}}{2}f_{2}(\psi)(Q \wedge \star Q) \\ \nonumber
+ \frac{\beta - \beta_{0}}{2}(Q \wedge \star Q) + \frac{\delta}{2}(d \psi \wedge \star d \psi) + F(e, \omega, ...)
\end{eqnarray}
Where $f_{1}(\psi)$ and $f_{2}(\psi)$ are 0-forms functions of the scalar field $\psi$ and $F(e, \omega,...)$ a generic $n$ form dependent on the coframe $e$, the connection $\omega$ but not on the scalar field $\psi$.\\
The Cartan equation can be written as:
\be
k \, D \star (e_{a} \wedge e^{b}) = F_{a}{}^{b}
\ee
Where $F_{a}{}^{b}$ are dependent on $e$ and $\omega$ and other fields but it is not dependent on $\psi$.\\
Suppose now that in the limit:
\begin{eqnarray}
f_{1}(\psi) \rightarrow 1 \\ \nonumber
f_{2}(\psi) \rightarrow 1
\end{eqnarray}
We get a Proca-type equation for $Q$
\be
\alpha d \star dQ + \beta_{0} \star Q = 0
\ee
and the action (1) in the limit (3) satisfies the Obukhov theorem, so that the generalised Einstein equations reduce to: \footnote{note that $F(e, \omega)$ is not supposed to vanish in the limit (3) since the effect of the reduction property is to 'cancel' the effect of the last three terms in the action (1).}
\be
k {\stackrel{o}{G}}_{c} + \tau_{c}[\alpha] + \tau_{c}[\beta_{0}] + \tau_{c}[\delta] = 0
\ee
where ${\stackrel{o}{G}}_{c}$ is the riemannian part of the Einstein $n-1$ forms ${\stackrel{o}{R}}^{a}{}_{b} \wedge \star (e_{a} \wedge e^{b} \wedge e_{c})$ and:
\begin{eqnarray}
\tau_{c}[\alpha] = \frac{\alpha}{2}(dQ \wedge i_{c} \star dQ - i_{c}dQ \wedge \star dQ) \\ \nonumber
\tau_{c}[\beta] = - \frac{\beta_{0}}{2}(Q \wedge i_{c} \star Q + i_{c} Q \wedge \star Q) \\ \nonumber
\tau_{c}[\delta] = -\frac{\delta}{2}(d \psi \wedge i_{c} \star d \psi + i_{c} d \psi \wedge \star d \psi)
\end{eqnarray}
\newpage
Then:\\
\\
{\bf Theorem 1}:\\
\\
\emph{The reduction occurs in the case $f_{1}(\psi) \neq 1, f_{2}(\psi) \neq 1$ as well, with (5) replaced by}
\be
k {\stackrel{o}{G}}_{c} + f_{1}(\psi) \tau_{c}[\alpha] + f_{2}(\psi) \tau_{c}[\beta_{0}] + \tau_{c}[\delta] = 0
\ee
\bigskip
{\bf Proof:}\\
\\
Doing the trace of the connection variation of (1) we get (mod d) \footnote{(mod d) is intended to mean that when we perform variations we may get also other terms which are total derivatives and then do not contribuite to the field equations.}:
\be
\alpha d(f_{1}(\psi) \star dQ) + f_{2}(\psi) \beta_{0} \star Q + (\beta - \beta_{0}) \star Q = \sum_{k} (\star E_{k})
\ee
In which $E_{k}$ are defined as follows:\\
From the variations of $\underbrace{F(e,\omega)}_{\omega} = \delta \omega^{a}{}_{b} \wedge \sum_{k} \star (B_{a}{}^{b})_{k}$ and then:
\be
2n \sum_{k} \star E_{k} = \sum_{k} \star (B_{a}{}^{a})_{k}
\ee
Now from the hypothesis we have for $f_{1}(\psi) = f_{2}(\psi) =1$ that:
\be
\alpha d \star dQ + \beta_{0} \star Q = 0
\ee
so we have:
\be
(\beta - \beta_{o}) \star Q = \sum_{k} (\star E_{k})
\ee
when we consider $f_{1}(\psi), f_{2}(\psi) \neq 1$ we are considering the replacement:
\begin{eqnarray}
\beta(Q	 \wedge \star Q) \rightarrow [\beta_{0} f_{2}(\psi) + (\beta - \beta_{0})(Q \wedge \star Q) \\ \nonumber
(dQ \wedge \star dQ) \rightarrow f_{1}(\psi) (dQ \wedge \star dQ)
\end{eqnarray}
Consider now the cartan equation which we may put in the form:
\be
k D \star (e_{a} \wedge e^{b}) = \delta^{b}{}_{a} A + F'^{b}{}_{a} 
\ee
by considering the trace we get:
\be
2n A + F'^{a}{}_{a} = 0
\ee
from which
\be
F'^{a}{}_{a} = - 2n A
\ee
so that we can rewrite (13) as:
\be
k D \star (e_{a} \wedge e^{b}) = -\delta^{a}{}_{b} \frac{F'^{a}{}_{a}}{2n} + F'^{b}{}_{a}
\ee
$F'^{a}{}_{b}$ is dependent on the fields $Q, T^{c}, \psi, ...$ so if the action is specified $F'^{b}{}_{a}$ and $F'^{a}{}_{a}$ are specified, the transformation (12) changes only $A$ in (13) and not $F'^{b}{}_{a}$ but relation (15) is still valid and so is (16). So Rel. (16) is not affected by the transformation (12). Then we obtain that the traceless part of the Cartan equation is invariant under the transformation (12). The consequence of this is that we can express $\lambda^{a}{}_{b}$ as a function of $F'^{b}{}_{a}$ and we will get the same expression for $\lambda^{a}{}_{b}$.\\
So we may formulate the following:\\
\\
{\bf Lemma}\\
\emph{The traceless part of the Cartan equation may be written in a form which is invariant under transformation like (12) of terms which contains only non-metricity and metric and connection invariant fields. The functional dependence of the non-Riemannian part of the connection is then unaffected.}\\
\\
By using the previous Lemma we can calculate the quantities $E_{k}$ appearing in (8) which will have the same functional dependence. \\
So we can conclude that the relation:
\be
(\beta - \beta_{0}) \star Q = \sum_{k} \star E_{k} 
\ee
is still valid since it may be expressed as a function of $\lambda^{a}{}_{b}$ and other fields which are independent on $\lambda^{a}{}_{b}$. \\
Then we get:
\be
\alpha \, d(f_{1}(\psi) \star dQ) + f_{2}(\psi) \beta_{0} \star Q = 0
\ee
with the same reasoning since the algebraic dependence of $\lambda^{a}{}_{b}$ on $F'^{a}{}_{b}$ is the same, we can state that the relation:
\be
k \Delta {\stackrel{o}{G}}_{c} + \tau_{c}[\beta - \beta_{o}] + \tau_{c}[F[e,\omega]] = 0
\ee
in which $\tau_{c}[F[e,\omega]]$ indicate the stress forms contribution due to $F(e,\omega)$ in the action (1), will be valid in general. \\
In the previous one $\Delta {\stackrel{o}{G}}_{c}$ is the non-Riemannian contribution to $G_{c}$. \\
Then we reduce the generalised Einstein equations to:
\be
k {\stackrel{o}{G}}_{c} + f_{1}(\psi) \tau_{c}[\alpha] + f_{2}(\psi) \tau_{c}[\beta_{0}] + \tau_{c}[\delta] = 0
\ee
\bigskip
so the theorem is proved.\\
\\
\\
The proof of the theorem and the Lemma relies essentially  on the fact the connection variation of terms like:
\begin{eqnarray}
f_{1}(\psi)(dQ \wedge \star dQ) \\ \nonumber
f_{2}(\psi)(Q \wedge \star Q)
\end{eqnarray}
contain the diagonal operator $\delta^{a}{}_{b}$.\\
It is clear that this property valid for scalar fields is valid for other metric and connection invariant fields.\\
Consider for example a term like:
\be
\star (F \wedge \star F) (dQ \wedge \star dQ)
\ee
where $F$ is the electromagnetic field $F = dA$, the connection variation is (mod d):
\be
2 \, \delta \omega^{a}{}_{b} \, \delta^{b}{}_{a} \wedge [d \star (F \wedge \star F) \wedge \star dQ]
\ee
The presence of the diagonal operator $\delta^{a}{}_{b}$ permits to extend what found to actions which contain terms analogous to (22). The equation for $Q$ will be modified but the reduction property is still valid with the proper stress forms originating from terms like (22) [8].\\
\\
\section{The Dilaton case}
\bigskip
In this section we extend the Obukhov theorem to a class of {\bf Dilaton 
non-Riemannian gravity}. We consider for simplicity the case in which the action contains only terms dependent on the non-metricity, that is we do not put any explicit torsion term.\\
We start the analysis from the action:
\be
S = \int [k \psi^2 R \star 1 + \beta (d \psi \wedge \star d \psi) - V(\psi) \star 1]
\ee
We use the Tucker-Wang gauge in which the metric is fixed to be orthonormal $g_{ab} = \eta_{ab} = (-1,1,1,1, ...)$, and we consider the coframe $e^{a}$ and the connection $\omega^{a}{}_{b}$ as independent variables.\\
As we will see the non-Riemannian contribution to the Einstein-Hilbert term times $\psi^2$ is equivalent in the field equations to a kinetic term for the dilaton.\\
The variation of (24) with respect to $\psi$ gives (mod d):
\be
-2(\beta d \star d \psi + k \psi (R \star 1)) - V'(\psi) \star 1 = 0
\ee
The variation with respect to the connection of action (24) gives:
\be
D \star (e_{a} \wedge e^{b}) = -\frac{2}{\psi}[d \psi \wedge \star (e_{a} \wedge e^{b})]
\ee
While the coframe variation gives:
\begin{eqnarray}
k \psi^2  {\stackrel{o}{R}}^{a}{}_{b} \wedge \star (e_{a} \wedge e^{b} \wedge e_{c}) - 
2 k \psi [{\hat{\lambda}}^{a}{}_{b} \wedge d \psi \wedge \star (e^{b} \wedge e_{a} \wedge e_{c})] \\ \nonumber
-\beta [d \psi \wedge i_{c} \star d \psi + i_{c} d \psi \wedge \star d \psi] 
+ k \psi^2[{\hat{\lambda}}^{a}{}_{d} \wedge {\hat{\lambda}}^{d}{}_{b}] \wedge \star (e_{a} \wedge e^{b} \wedge e_{c}) - \\ \nonumber
 V(\psi) \star e_{c} = 0
\end{eqnarray}
where use has been made of the expression for the full non-Riemannian Einstein-Hilbert term:
\be
R \star 1 = {\stackrel{o}{R}} \star 1 - {\hat{\lambda}}^{a}{}_{c} \wedge {\hat{\lambda}}^{c}{}_{b} \wedge \star (e^{b} \wedge e_{a}) - d({\hat{\lambda}}^{a}{}_{b} \wedge \star (e^{b} \wedge e_{a}))
\ee
The solution of the Cartan equation is:
\be
\lambda_{ab} = - \frac{1}{2n} g_{ab}Q + \frac{2}{(2-n) \psi}(i_{a}(d \psi) e_{b} - i_{b}(d \psi) e_{a})
\ee
and the traceless part is:
\be
{\hat{\lambda}}_{ab} = \frac{2}{2-n} \frac{1}{\psi} (i_{a}(d \psi) e_{b} - i_{b} (d \psi) e_{a})
\ee
by using this expression in the generalised Einstein equations we get:
\be
k \psi^2 {\stackrel{o}{G}}_{c} - \beta' [d \psi \wedge i_{c} \star d \psi + i_{c} d \psi \wedge \star d \psi] - V(\psi) \star e_{c} = 0
\ee
where:
\be
\beta' = \beta + 4k \frac{n-1}{n-2}
\ee
So the presence of the full non-Riemannian Einstein-Hilbert term is equivalent to a rescaling of the scalar kinetic term coupling constant.\\
Observe the following two interesting cases:\\
\\
a] If we have $\beta = 0$ in the action (24) then we get the generalised Einstein equations:
\be
k \, \psi^2 {\stackrel{o}{G}}_{c} + 4k \frac{n-1}{n-2} [d \psi \wedge i_{c} \star d \psi + i_{c} d \psi \wedge \star d \psi] - V(\psi) \star e_{c} = 0
\ee
then for $n = 4$ 
\be
k \psi^2 {\stackrel{o}{G}}_{c} + 6 \, k [d \psi \wedge i_{c} \star d \psi + i_{c} d \psi \wedge \star d \psi] - V(\psi) \star e_{c} = 0
\ee
So we get the Dilaton Einstein equation with the conformal coupling $\xi = \frac{1}{6}$\\
\\
\\
b] If $\beta + 4k \frac{n-1}{n-2} = \beta' = 0$ or $\beta = - 4k \frac{n-1}{n-2}$.\\
\\
Then we get:
\be
k \psi^2 {\stackrel{o}{G}}_{c} - V(\psi) \star e_{c} = 0
\ee
which are equivalent to:
\be
k {\stackrel{o}{G}}_{c} - \tilde{V}(\psi) \star e_{c} = 0
\ee
where $\tilde{V}(\psi) = \frac{V(\psi)}{\psi^2}$. So the generalised Einstein equations are equivalent to the decoupled case with the potential $\tilde{V}(\psi)$.\\
We have to observe that the equations for the scalar field are different from the Levi-Civita one, so the reduction property refers only to the Einstein sector of the theory.\\
Let us consider now the action:
\be
S = \int [k \psi^2 R \star 1 + \beta (d \psi \wedge \star d \psi) + \frac{\alpha}{2}f_{1}(\psi)(dQ \wedge \star dQ) + \frac{\gamma}{2} f_{2}(\psi)(Q \wedge \star Q) - V(\psi) \star 1]
\ee
Using the Lemma of previous section we may certainly formulate the following:\\
\\
{\bf Theorem 2:}\\
\\
\emph{Suppose we start from the action:}
\be
S = \int [k \psi^2 R \star 1 + \beta (d \psi \wedge \star d \psi) + f_{1}(\psi) \frac{\alpha}{2}(dQ \wedge \star dQ) + f_{2}(\psi) \frac{\gamma}{2}(Q \wedge \star Q)] - \\ \nonumber
V(\psi) \star 1
\ee
\emph{Then the generalised Einstein equations can be reduced to:}
\begin{eqnarray}
k \psi^2 {\stackrel{o}{G}}_{c} - \beta' [d \psi \wedge i_{c} \star d \psi + i_{c} d \psi \wedge \star d \psi] - V(\psi) \star e_{c} - \\ \nonumber
+ f_{1}(\psi) \frac{\alpha}{2}[dQ \wedge i_{c} \star dQ + i_{c} dQ \wedge \star dQ] + f_{2}(\psi) \frac{\gamma}{2}[Q \wedge i_{c} \star Q - i_{c}Q \wedge \star Q] = 0
\end{eqnarray}
\emph{with} $\beta' = \beta + 4k \frac{n-1}{n-2}$.\\
\\
------------------\\
\\
Eq (39) can be considered the generalization of the Obukhov-Tucker-Wang theorem to the non-Riemannian Dilaton Gravity action (37) [9].\\
The equation for $\psi$ becomes:
\be
- 2(\beta d \star d \psi + k \psi (R \star 1)) - V'(\psi) \star 1 + f_{1}'(\psi) \frac{\alpha}{2}(dQ \wedge \star dQ) + f_{2}'(\psi) \frac{\gamma}{2} (Q \wedge \star Q) = 0
\ee
From the trace of the Cartan equation we get the new equation for $Q$:
\be
\alpha \, d(f_{1}(\psi) \star dQ) + \gamma f_{2}(\psi) \star Q = 0
\ee
In the action (38) we have not introduced terms like $T \wedge \star T$, $T^{c} \wedge \star T_{c}$, ($T$ is defined as $i_{c}T^{c}$) . \\
Had we done that we would have modified the traceless part of the Cartan equation so the Lemma proved in section 2 which brought us to Theorems 1,2 would not be valid anymore. We expect however a similar result to hold for an action more general than (38) but the proof is not trivial and may require a computer based calculation, so much said, we hope to obtain this further generalization in the next paper of this series.\\
\\
{\bf Acknowledgments}\\
\\
The NOOPOLIS foundation (Italy) is acknowledged for partial funding, also the discussions with R. Tucker and C. Wang were very fruitful.

\newpage
\begin{center}
{\bf REFERENCES}
\end{center}
\bigskip
1] F. W. Hehl, J. D. McCrea, E. W. Mielke, Y. Neeman, Phys. Rep. {\bf 258} 1 (1995), and F. Gronwald, Int. J. Mod. Phys. D, {\bf 6} 263 (1997).\\
\\
2] R. W. Tucker, C. Wang, Class. Quant. Grav. {\bf 12} 2587 (1995).\\
\\
3] R. Tucker. C. Wang, Non-Riemannian Gravitational Interactions, Institute of Mathematics, Banach Center Publications, Vol. 41, Warzawa (1997).\\
\\
4] T. Dereli, M. Onder, J. Schray. R. Tucker, C. Wang, Class. Quant. Grav. {\bf 13} L 103 (1996) .\\
\\
5] Y. Obukhov, E. Vlachynsky, W. Esser, F. Hehl, Phys. Rev. D {\bf 56} 12, 776 (1997).\\
\\
6] F. W. Hehl, A. Macias, gr-qc/9902076, Int. J. Mod. Phys. D, (1999) (to appear).\\
\\
7] R. Scipioni, Class. Quant. Grav. {\bf 16} 2471 (1999).\\
\\
8] R. Scipioni, PhD thesis (1999) Lancaster.\\
\\
9] R. Scipioni, Phys. Lett. A (in press).\\
\\
\end{document}